\documentclass[aps,prb,twocolumn,superscriptaddress,amsmath,amssymb,showpacs]{revtex4}
\usepackage{graphicx}
\usepackage{dcolumn}
\usepackage{bm}
\begin{document}

\title{Dielectric quantification of conductivity limitations due to nanofiller size in conductive
powders and nanocomposites}

 \author{L. J. Huijbregts}
 \affiliation{Technische Universiteit Eindhoven, POB. 513, 5600 MB Eindhoven, The Netherlands}
 \affiliation{Dutch Polymer Institute (DPI), POB. 902, 5600 AX Eindhoven,The Netherlands}
 \author{H. B. Brom}
 \affiliation{Technische Universiteit Eindhoven, POB. 513, 5600 MB Eindhoven, The Netherlands}
 \affiliation{Dutch Polymer Institute (DPI), POB. 902, 5600 AX Eindhoven,The Netherlands}
 \affiliation{Kamerlingh Onnes Laboratory, Leiden University, POB. 9504, 2300 RA Leiden, The Netherlands}
 \author{J. C. M. Brokken-Zijp}
 \affiliation{Technische Universiteit Eindhoven, POB. 513, 5600 MB Eindhoven, The Netherlands}
 \affiliation{Dutch Polymer Institute (DPI), POB. 902, 5600 AX Eindhoven,The Netherlands}
 \author{W. E. Kleinjan}
 \affiliation{Technische Universiteit Eindhoven, POB. 513, 5600 MB Eindhoven, The Netherlands}
 \author{M. A. J. Michels}
 \affiliation{Technische Universiteit Eindhoven, POB. 513, 5600 MB Eindhoven, The Netherlands}
 \affiliation{Dutch Polymer Institute (DPI), POB. 902, 5600 AX Eindhoven,The Netherlands}

 \date{December 28 2007 accepted for Phys. Rev. B}

 \begin{abstract}
Conducting submicron particles are well-suited as filler particles
in non-conducting polymer matrices to obtain a conducting
composite with a low percolation threshold. Going to
nanometer-sized filler particles imposes a restriction to the
conductivity of the composite, due to the reduction of the density
of states involved in the hopping process between the particles,
compared to its value within the crystallites. We show how those
microscopic parameters that govern the charge-transport processes
across many decades of length scales, can accurately and
consistently be determined by a range of dielectric-spectroscopy
techniques from a few Hz to infrared frequencies. The method,
which is suited for a variety of systems with restricted
geometries, is applied to densely packed 7-nm-sized tin-oxide
crystalline particles with various degree of antimony doping and
the quantitative results unambiguously show the role of the
nanocrystal charging energy in limiting the hopping process.
\end{abstract}


 \maketitle

\section{Introduction}

Small submicron particles are well-suited as fillers in
non-conducting polymer matrices to obtain a conducting composite
with a percolation threshold (far) below 1 \%. The low percolation
threshold is due to the formation of airy aggregates of conducting
particles, in which the particles are grown together by
diffusion-limited cluster aggregation, creating a network with a
fractal dimension around 1.7.
\cite{BundeHavlin91,VDPuttenAdriaanse9297,Grimaldi0306} These airy
aggregates can be thought of as conducting spheres forming a
3-dimensional percolating network around the expected aggregate
filling fraction 0.16. As a consequence of the fractal structure
within the aggregates, the filler fraction of the particles at the
percolation point is much lower. Even in case the particles touch,
the dc conductivity ($\sigma_{\rm dc}$) of these composites at
high filling fractions turns out to  be orders of magnitude lower
than of the bulk material, as was recently illustrated for a
particular crosslinked epoxy composite with filler particles of
Phthalcon-11, \cite{BrokkenZijp95} Co phthalocyanine crystallites
of 100 nm size, and explained by purely structural
arguments.\cite{Huijbregts06JPC}

When crystalline particles with a diameter of less than 10 nm
instead of 100 nm are used,  the small size of the particles may
impose another important restriction to the maximal possible
composite conductivity, which is due to the density of states
(DOS) involved in the dc conductivity through the network of
particle contacts. Compared to larger crystallites, this DOS can
be strongly reduced by the charging
energy.\cite{Yu04,Zhang04,Beloborodov05,Feigelman05}

We show how those microscopic parameters, which govern the
charge-transport process across many decades of length scales, can
accurately and consistently be determined by ac (alternating
current) dielectric spectroscopy from a few Hz to infrared
frequencies. In particular we can address the parameters for Mott
variable-range hopping, for heterogeneity-induced enhanced ac
response, for phonon- or photon-assisted nearest-neighbor hopping,
and for the Drude response of individual nanocrystals. Due to
these quantitative results we can unambiguously determine also the
role of the nanocrystal charging energy in limiting the hopping
process. We apply the method to antimony-doped tin-oxide (ATO)
crystallites of 7 nm diameter and to 100 nm sized crystallites of
Phthalcon-11. It turns out that in densely packed crystallites of
ATO, due to the strong influence of the charging energy on the
DOS, $\sigma_{\rm dc}$ at room temperature is four orders of
magnitude lower than the dc conductivity extrapolated from the
Drude plasma frequency ($\omega_{\rm pD}$) of the crystallites - a
result with obvious implications for the design of conducting
composites. The dielectric method is well suited for a variety of
systems with restricted geometries, as we will illustrate by a
short discussion of phase-change materials \cite{WelnicWuttig07}
and granular oxides. \cite{BenChorin93Vaknin00Orlyanchik07}.

\section{Conductivity in conducting polymer composites and granular metals}

For randomly placed conducting spheres in an insulating matrix,
the relation between $\sigma_{\rm dc}$ and the fraction $\phi$ of
spheres is known from percolation theory.
\cite{Stauffer85,BundeHavlin91,Straley7677}Above the percolation
threshold $\sigma=\sigma_0|\phi-\phi_c|^t$ where $\phi_c \approx
0.16$ is the percolation threshold, $t \approx 2.0$ and $\sigma_0$
is approximately equal to the conductivity of the spheres.
\cite{Zallen83,Grimaldi0306}  When the building blocks of the
network are fractal aggregates instead of solid spheres,
$\sigma_0$ has to be replaced by the aggregate conductivity
$\sigma_{\rm a}$ and depends on the particle conductivity and, via
the non-linear relation  $\sigma_{\rm a}/\sigma_{\rm p} =
(\phi_{\rm p,c} /0.16)^{1+x}$ with $x>0$, on the real percolation
threshold $\phi_{\rm p,c}$ of the particles.\cite{Huijbregts06JPC}
The value of the exponent $x$ is related to the random-walk
dimension and the fractal dimension $d_f$, and is maximally
$2(d_f-1)/(3-d_f)$. This shows that on purely geometrical grounds
for a network with $\phi_{\rm p,c}=0.0055$, at the highest filling
fraction of the aggregates $\sigma_{\rm dc}$ will be three to four
orders of magnitude lower than in the pure filler powder.
\cite{Huijbregts06JPC}

As remarked in the introduction, when nanosized particles are used
as fillers, charging energies (and quantum size effects) impose a
further important restriction to the maximal possible composite
conductivity.\cite{Yu04,Zhang04}  This effect can be conveniently
studied in densely packed powders of filler material by dielectric
spectroscopy.

\subsection{Parameters for the dc conductivity}
In the ohmic regime, if there is a non-negligible density of
states around the chemical potential, and the temperature $T$ is
high enough that also the Coulomb interaction can be neglected
($T>T_{\rm crit}$), $\sigma_{\rm dc}$ will obey Mott's equation
for conduction via variable-range hopping (VRH): \cite{Mott69}
 \begin{equation}
 \sigma_{\rm dc} \propto \exp[-(T_{\rm 0,Mott}/T)^{\nu}]
 \label{esM}
 \end{equation}
 with $\nu=1/4$ and
 \begin{equation}
 k_BT_{\rm 0,Mott} \approx 20 /(g_{\rm hop} a^3),
 \label{eT0M}
 \end{equation}
where $a$ denotes the decay length of the electron density,
$k_{\rm B}$ the Boltzmann constant, and $g_{\rm hop}$ the density
of states relevant in the hopping process. \cite{Bottger85}

For randomly packed spheres of radius $R$ and spacing $s$ the
localization length  $\tilde{a}$ will be enlarged,
\cite{Nemeth88,Zvyagin99,Zhang04,Beloborodov05,Feigelman05} and
can be approximated by \cite{Nemeth88,Zhang04}
\begin{equation}
\tilde{a} = (2R/s)a.
 \label{ea}
\end{equation}
In the following we drop the tilde.

Below $T_{\rm crit}$ the $T$ dependence of the conductivity will
be dominated by a soft Coulomb gap, leading to so-called
Efros-Shklovskii (ES) VRH: \cite{Shklovskii84}
 \begin{equation}
 \sigma_{\rm dc} \propto \exp[-(T_{0,\rm ES}/T)^{1/2}].
 \label{esES}
 \end{equation}
In the ES VRH model in the dilute limit of a large distance
between the particles $T_{0,\rm ES}$ is given by
 \begin{equation}
 T_{0,\rm ES} = 2.8 e^2 /(4 \pi \epsilon \epsilon_0 a k_{\rm B}),
 \label{eT0ES}
 \end{equation}
with $e$ the electron charge, $\epsilon_0$ the vacuum dielectric
constant, and $\epsilon$ the relative dielectric constant of the
medium. $T_{\rm crit}$ is given by $ T_{\rm crit} \sim e^4 a
g_{\rm hop} / [k_{\rm B} (4 \pi \epsilon \epsilon_0)^2]$ and (for
$s>R$) the charging energy by $\sim e^2/(4 \pi \epsilon \epsilon_0
R)$.

For densely packed small particles, at high temperatures but still
in the regime, where Coulomb interactions are important ($T\leq
T_{\rm crit}$), ES VRH behavior will evolve into nearest-neighbor
hopping at a temperature $T_{\rm A}$. Above $T_{\rm A}$ the
conduction is thermally activated with an activation energy
$\Delta E_{\rm A}$ of the order of the charging energy, and
$T_{\rm A}\sim (\Delta E_{\rm A})^2/T_{\rm 0,ES}$.

The experiments of Yu {\it et al.} \cite{Yu04,noteCdSe1} on thin
films of highly monodispersed semiconducting nanocrystals of CdSe
of 6 nm diameter, slightly smaller than the ATO crystallites
discussed here, showed good agreement between the theoretical and
experimental value of $T_{0,\rm ES}$, $T_{\rm crit}$ and $T_{\rm
A}$. \cite{noteCdSe2}

\subsection{Sub-THz and far-infrared regime}
At sufficiently low frequencies the conductivity will be frequency
independent and equal to its dc value, because the inhomogeneities
are averaged out by the motion of the charge carriers. The minimal
length scale for homogeneity is referred to as $L_{\rm hom}$,
 \begin{equation}
 L_{\rm hom}^2 = \sigma_{\rm dc} k_B T / n_{\rm hom} e^2 (f_{\rm
 os}/2D),
 \label{eLhom}
 \end{equation}
where  $n_{\rm hom}$ is the density of the carriers involved in
the hopping process at the border of the homogenous regime, and
the onset frequency $f_{\rm os}$ for the frequency dependence of
$\sigma$ is divided by $2D$, with $D$ the dimension of the system.
\cite{Bottger85}

At high enough frequencies, when during half a period of the
oscillation of the applied field electrons can hop solely between
nearest-neighbors, the major contribution to the conductivity will
be due to tunnelling between localized states at neighboring sites
(the pair limit).\cite{Pollak61} This incoherent process can be
either by phonon-assisted or photon-assisted hopping, where in the
latter case the energy difference between the sites is supplied by
photons instead of phonons. \cite{Bottger85} The phonon-assisted
contribution to the conductivity is given by
 \begin{equation}
 \sigma_{\rm phonon}(\omega) = \frac{\pi^2}{192} e^2\omega k_{\rm
B}
 T L_{\rm hop}^5g_{\rm hop}^2\ln^4 \left({\frac{\omega_{\rm ph}}{\omega}}\right),
 \label{esphon}
 \end{equation}
with $L_{\rm hop} \sim a$ the decay length of the electronic state
outside the conducting particles, $g_{\rm hop}$ the relevant DOS
at the Fermi energy $E_{\rm F}$, and $\omega_{\rm ph}$ the phonon
`attempt' frequency.\cite{Bottger85} This formula is valid when
$\omega < \omega_{\rm ph}$; at higher $\omega$, where the
contribution of phonon-assisted hopping to $\sigma$ becomes
constant, photon-assisted processes usually take over, with a
conductivity $\sigma_{\rm photon}$ given by
 \begin{equation}
 \sigma_{\rm photon}(\omega) = \frac{\pi^2}{6}
 e^2\hbar\omega^2 L_{\rm hop}^5g_{\rm
hop}^2\ln^4{\left(\frac{2I_0}{\hbar\omega}\right)}.
 \label{esphot}
 \end{equation}
The energy $k_{\rm B}T$ in eq.~(\ref{esphon}) is in
eq.~(\ref{esphot}) replaced by $\hbar\omega$ and the phonon
attempt frequency $\omega_{\rm ph}$ by $2I_0$, with $I_0$ being
the `overlap' pre-factor for the energy levels of two neighboring
sites. In analogy with $\omega_{\rm ph}$, $I_0/\hbar$ can be
interpreted as the attempt frequency for photon-assisted hopping.
Equation (\ref{esphot}) is only valid when $\omega < I_0/\hbar$.
As in phonon-assisted hopping, $\sigma$ passes over into a plateau
at high $\omega$.

\subsection{Visible and optical response}
At high frequencies (for ATO in the infrared regime) the short
period of the electromagnetic field will restrict the motion of
the carriers to the nanocrystallite, and the dielectric response
characterized by the complex relative dielectric constant
$\epsilon(\omega) = 1 - \omega_{\rm pD}^2 /
[\omega(\omega+i\Gamma)]$ will be Drude-like, with $\omega_{\rm
pD}$ the Drude plasma resonance frequency  and $\Gamma=1/\tau$ the
damping rate. In practice the constant 1 has to be replaced by
$\epsilon_{\infty}$ due to other contributions in this frequency
regime, like the polarization of the ion cores. \cite{Kittel96}
The Drude plasma frequency is related to the number of carriers
per unit of volume $n_{\rm crys}$ and the effective mass $m_{\rm
e}^*$ as
 \begin{equation}
 \omega_{\rm pD}^2 = n_{\rm crys} e^2 / \epsilon_0
 m_{\rm e}^*.
 \label{epD}
 \end{equation}
For damping rates comparable to the Drude plasma frequency, the
real plasma frequency (where the dielectric constant becomes zero)
will be larger than $\omega_{\rm pD}$. $\Gamma$ is determined by
the boundaries of the nanoparticle and additional (ionized
impurity) scattering:
 \begin{equation}
 \Gamma=1/\tau=v_{\rm F}/L_{\rm crys},
 \label{egamma}
  \end{equation}
where $1/L_{\rm crys}$ is the sum of the inverse size of the
particle and the inverse phonon scattering length.

\section{Experimental procedures and data}

Measurements were performed on Sb-doped tin-oxide nanoparticles
with [Sb]/([Sn]+[Sb]) equal to 0, 2, 5, 7, 9, and 13 at.\%. The
particles are monocrystalline and spherical with diameters close
to 7 nm. \cite{Kleinjan08} The diameter of the 7\% doped
crystallite is 7.1 nm. Sb is incorporated in the casserite SnO$_2$
lattice by replacing Sn$^{4+}$. At the doping level of 7\%,  Sb is
mainly present as Sb$^{5+}$, resulting in n-type conductivity of
the ATO particles according to Nutz {\it et al.}. \cite{Nutz99}
The amount of Sb$^{3+}$ present in the particles is negligibly
small. \cite{Kleinjan08,McGinley01}

The followed experimental procedures for the dc conductivity and
dielectric measurements are described in Refs.
\onlinecite{Huijbregts06JPC} and \onlinecite{Huijbregts0608PSS}.
The thickness of the samples was typically a few mm. The dc
conductivity measurements were performed in the dark under helium
atmosphere.

\begin{figure}[htb]
 \begin{center}
 \includegraphics[width=8cm]{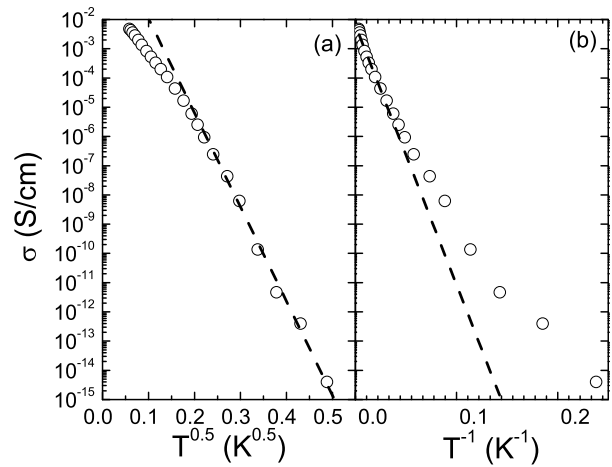}
 \end{center}
\caption{dc conductivity as function of $T$ for a densely packed
powder of  7\%-doped ATO. For $T \leq  50$~K the data can be
fitted by eq.~(\ref{esES}) (a), while for $T \geq  50$~K the $T$
dependence is activated (b).}\label{svsT}
\end{figure}

\begin{figure}[htb]
 \begin{center}
 \includegraphics[width=8cm]{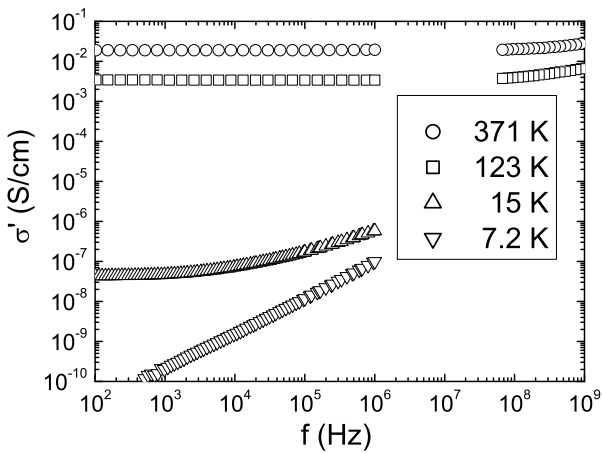}
 \end{center}
\caption{Frequency dependence of the conductivity for 7\%-doped
ATO at various temperatures. In this double logarithmic plot the
linear dependence at low temperatures is in agreement with
eq.~(\ref{esphon}).}\label{svsf}
\end{figure}

The $T$ dependence is given in Fig.~\ref{svsT}, and the frequency
dependence at temperatures down to 7 K in Fig. \ref{svsf}. All
data shown are for 7\%-doped ATO. Similar results were obtained at
other doping levels, be it with different absolute values. The
data were taken in the ohmic regime.

\begin{table}[htb]
\vfill
\begin{center}
 \begin{tabular}{|c|c|c|c|c|c|c|} \hline
 sample                  & N             &$\sigma_{\rm dc}$&$\omega_{\rm pD}$&$\Gamma$         & $m_{\rm e}^*$  \\ \hline
                         & (cm$^{-3}$)   & (S/cm)          & (s$^{-1}$)      & (s$^{-1}$)      & ($m_{\rm e}$) \\ \hline
 ATO-film\cite{Shanthi80}&$2\cdot10^{20}$&$10^{2}$         &$10^{15}$        &$8\cdot10^{14}$  & 0.3  \\ \hline
 ITO-film\cite{Mergel02} &$10^{21}$      &                  &$5\cdot10^{14}$  &$3\cdot10^{14}$  & 0.35 \\ \hline
 ATO-powder\cite{Nutz99} &$2\cdot10^{21}$&$10^{-6}$        &$10^{15}$        &$6\cdot10^{14}$  & 0.27 \\ \hline
 ATO-powder(pw)          &$10^{21}      $&$10^{-2}$        &$4\cdot10^{14}$  &$10^{14}$        & 0.3 \\ \hline
 \end{tabular}
\caption{Transport parameters obtained for ATO and indium tin
oxide (ITO) (second row) at doping levels of $10^{20}$ to
$10^{21}$ per cm$^3$. The room-temperature dc conductivities are
given in S/cm, the Drude frequencies and damping rate in s$^{-1}$,
and the effective mass in free-electron masses. The first and
second row are obtained for films of ATO \cite{Shanthi80} and ITO
\cite{Mergel02} resp., the last two rows contain the data on
powders of 6\% doped ATO particles of N\"{u}tz {\it et
al.}\cite{Nutz99} and our data (labelled as pw for present work)
on samples with 7\% Sb doping. For a bulk material with a Drude
frequency of $10^{15}$ Hz, a scattering time of $10^{-14}$~s and a
carrier mass of 0.3 $m_{\rm e}$ a dc conductivity is expected of
$10^2$ S/cm } \label{numbers}
\end{center}
\end{table}

\begin{figure}[htb]
\begin{center}
 \includegraphics[width=8cm]{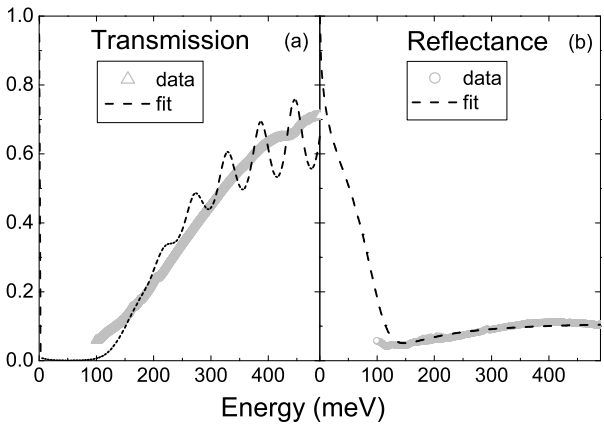}
\end{center}
\caption{Transmittance and reflectance of ATO with 7\% Sb doping
in the infrared, after background correction. (a) For the
transmission ATO is mixed with KBr and pressed into a 0.57 mm
thick pellet to have sufficient transparency. The oscillations in
the fit to the transmittance are an artefact (the fit is based on
the calculated effective ATO-film thickness of 0.005 mm). (b) Like
for the sub-THz data, the reflection data are taken from
precipitated films of 1 mm thick typically. The Drude fit is
discussed in the text. The data compare well with the data
published in the literature, see Table \ref{numbers}.}
\label{TRIR}
\end{figure}

The infrared (IR) transmittance was measured on a pellet of KBr
mixed with a small amount of ATO. For the IR-reflectance we used a
precipitated film of ATO with a thickness of about 1 mm. The data
are shown in Fig.~\ref{TRIR}.  For the analysis we also used the
sub-THz transmittance and phase data (only shown in
Fig.~\ref{sfit}).

In Table \ref{numbers} we summarize our data on densely packed
7-nm-sized ATO crystallites and compare them with measurements on
doped tin oxide published in the literature. The values of
$\omega_{\rm pD}$ agree within a factor 2, while the spread in the
scattering rates is larger.

\section{Analysis and Discussion}

In the analysis we first show the procedure to extract the
parameter values from the data in the different frequency regimes
and to check their consistency. We also make a comparison to the
parameter values of Phthalcon-11, for which the data are published
elsewhere. \cite{Huijbregts0608PSS} Then we concentrate on the
density of states; the latter being important for the dc
conductivity. Subsequently, the implications for the use of the
particles as fillers in nanocomposites are discussed.

\subsection{Procedure}
Regarding the $T$ dependence of $\sigma$ (Fig.~\ref{svsT}), the
data can be fitted with $\nu=1$ (eq.~(\ref{esM})) if the fit is
restricted to $T \geq 50$~K and with $\nu=0.5$ (eq.~(\ref{esES}))
for $T \leq 50$~K. The $\nu=1$ fit gives an activation energy of
$10^2$~K, while the exponent $\nu=0.5$ at low $T$ gives $T_{0,\rm
ES}=0.6 \cdot 10^4$~K.  The localization length from $T_{0,\rm
ES}$, see eq.~(\ref{eT0ES}), is calculated to be $a$ = 3 nm. Using
eq.~(\ref{ea}) and $(2R/s)= 10$ estimated from the packing
density, we find $a \sim 2$~nm, in good agreement with the value
calculated from $T_{0,\rm ES}$.

The onset of the frequency dependence of the conductivity (see
Fig.~\ref{svsf}) signals that the carrier starts to feel the
inhomogeneity of the underlying structure. Using eq.
(\ref{esphon}), the typical length scale $L_{\rm hom}$ at the
onset can be found. For 7\%-doped ATO at 300~K, the onset
frequency $f_{\rm os} = 3 \cdot 10^8$~Hz and $\sigma_{\rm dc} =
10^{-2}$~S/cm give a value of $L_{\rm hom}^2 n_{\rm hom}= 9 \cdot
10^7$ cm$^{-1}$.

\begin{figure}[htb]
\begin{center}
 \includegraphics[width=8cm]{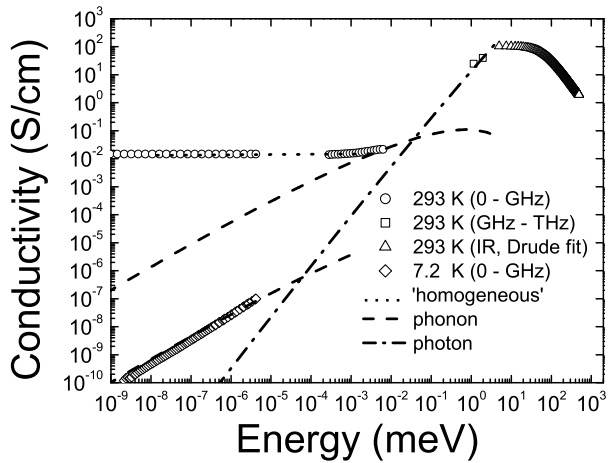}
\end{center}
\caption{The various contributions to the conductivity as function
of photon energy. At room temperature at the lowest frequencies
the conductivity is dominated by charging energies, at
intermediate frequencies phonon and photon assisted hopping
processes describe the frequency dependence, and in the infrared
the Drude conductivity inside the crystallites is seen.}
\label{sfit}
\end{figure}

The linear frequency dependence of the conductivity at 7~K in the
double logarithmic plot of Fig. \ref{svsf}, is in agreement with
phonon-assisted tunneling, see eq.~(\ref{esphon}). In the range of
10 - 100 cm$^{-1}$ photon-assisted processes take
over.\cite{Reedijk9899} Applying eq. (\ref{esphon}) to the
conductivity data at 293 K and taking the usual value for the
phonon frequency in solids $\omega_{\rm ph}=10^{12}$
s$^{-1}$,\cite{Bottger85,Reedijk9899} we find $L_{\rm hop}^5g_{\rm
hop}^2 = 3 \cdot 10^3$ eV$^{-2}$cm$^{-1}$, see Fig. \ref{sfit}.

Turning next to the high-frequency data presented in
Fig.~\ref{TRIR}, we performed a simple Drude analysis. The fit
($\omega_{\rm pD}$ = 11000 cm$^{-1}$ and $\tau$ = 3300 cm$^{-1}$,
together with a dielectric constant of 4.0) reproduces the main
features of the increase of the transmission and the level of the
reflectance (the oscillations in the fit to the transmittance are
an artefact because the effective ATO film thickness of 0.005 mm
is much smaller than the real thickness of the pressed KBr
pellet). The number of carriers $n_{\rm crys}$ of $10^{21}$
cm$^{-3}$ is directly derived from the Drude frequency and is
slightly lower than obtained from a simple interpretation of the
chemical composition. The bulk dc conductivity calculated from the
Drude parameters is $10^2$ S/cm. The fit parameters of the present
samples are given in Table \ref{numbers} and agree well with the
literature.

Fig.~\ref{sfit} shows the reconstructed conductivity of ATO as
function of frequency due to the processes discussed above.

For ATO the important values for the dc conductivity can be
deduced from the combination of variables that we found from the
previous analysis (i) $L_{\rm hom}^2 n_{\rm hom}= 9 \cdot 10^7$
cm$^{-1}$, (ii) $g_{\rm hop}^2 L_{\rm hop}^5 = 8 \cdot 10^{43}$
J$^{-2}$m$^{-1}$ or $3 \cdot 10^3$ ${\rm eV^{-2}cm^{-1}}$ for
photon-assisted hopping and $10^{42}$ J$^{-2}$m$^{-1}$ for the
phonon-fit to the data at 7 K, and (iii) $n_{\rm crys} = 10^{21}$
cm$^{-3}$ and $\tau=10^{-14}$ s.

Using (iii) the `extrapolated' dc conductivity is $10^2$ S/cm, a
factor $10^4$ larger than the found value of $10^{-2}$~S/cm. The
estimated Fermi energy $E_{\rm F}$ is around 2 eV, and from
$g(E)=(3/2)(n / E_{\rm F})$ (valid for free electrons) we get
$g(E_{\rm F}) = 5 \cdot 10^{21}$ eV$^{-1}$cm$^{-3}$. From (ii)
with $L_{\rm hop} = a$ of 3 nm, we find for $g_{\rm hop}= 3 \cdot
10^{18}$ eV$^{-1}$cm$^{-3}$, a factor $10^3$ lower than $g(E_{\rm
F})$. Note that this is an averaged density of states involved in
photon assisted hopping. Due to the curvature of the density of
states around the chemical potential, $g(E)$ will be lower at
lower energies. For example, for the phonon-fit at 7~K $g_{\rm
hop}$ is equal to $0.3 \cdot 10^{18}$ eV$^{-1}$cm$^{-3}$.

The values for $\tau$ and $L_{\rm hom}$ can be used as a
consistency check. The combination of the estimated Fermi velocity
of $0.8 \cdot 10^8$~cm/s, with the crystallite size of 7 nm and
$m_{\rm e}^* =  0.3 m_{\rm e}$,\cite{Nutz99} predicts a surface
scattering rate of $10^{14}$~s$^{-1}$, in agreement with the found
value of $\Gamma$. Next, from $g_{\rm hop} k_B T \sim n_{\rm hom}$
eV$^{-1}$cm$^{-3}$ we now can estimate $n_{\rm hom}$ at $T =
293$~K as $n_{\rm hom} = 10^{17}$ cm$^{-3}$. Using (i) and $n_{\rm
hom}$ we find $L_{\rm hom}= $ 0.3 $\mu$m.

In short, the dielectric data of ATO allow a consistent picture of
the conduction process. In these densely packed crystallites the
localization length is enhanced by a factor 10 and the density of
states involved in the dc conductivity is more than a factor
$10^3$ smaller than that in the conduction within the
crystallites. The relatively large length scale for homogeneity is
indicative for the presence of aggregates. Indeed, like in
Ketjen-Black, \cite{VDPuttenAdriaanse9297} nanoparticles of ATO
are known to form chemically bonded aggregates that survive the
preparation stage.\cite{Kleinjan08} Due to the nature of the
chemical bond, the conductivities between neighboring crystallites
in and outside the aggregates are expected to be only slightly
different. Note that also the value of $g_{\rm crys}$ has to be
seen as an average, as inhomogeneities in the doping of ATO might
be present as well.\cite{McGinley01,Kleinjan08}

For the studied Phthalcon-11 crystallites $\omega_{\rm pD} =
10^{13}$ s$^{-1}$ and $\tau = 10^{-13}$ s leading to $n_{\rm crys}
= 2 \cdot 10^{15}$ cm$^{-3}$, i.e. about 1 charge per
crystallite.\cite{Huijbregts0608PSS} The other values found for
Phthalcon-11 are: (i) $L_{\rm hom}^2n_{\rm hom} = 2 \cdot 10^6$
cm$^{-1}$, (ii) $g_{\rm hop}^2 L_{\rm hop}^5 = 10^{43}$
J$^{-2}$m$^{-1}$ or $2.5 \cdot 10^3$ ${\rm eV^{-2}cm^{-1}}$.  In
these organic crystals with such a low carrier density, the charge
carriers can be seen as an electron gas with an energy scale set
by $k_{\rm B}T$, and $g(E)$ can be estimated from $g(E)k_{\rm B}T
\sim n_{\rm crys}$ to be $10^{17}$ ${\rm eV^{-1}cm^{-3}}$. This
value of $g(E)$ is the upper limit for $g_{\rm hop}$ and $g_{\rm
hom}$. From $g_{\rm hom}=10^{17}$ ${\rm eV^{-1}cm^{-3}}$, we find
a decay length $a$ of 3 nm, as expected from the packing.

The Phthalcon-11 parameters show that the crystals are
semiconducting crystals with a low number of charge carriers. All
charges participating in the conductivity within the crystal also
contribute to the dc conductivity. As for ATO the obtained
conduction parameters for Phthalcon-11 from the dielectric scans
give a consistent picture.

\subsection{Density of states}

For ATO, the differences between the density of states involved in
the hopping process $g_{\rm hop}= 3 \cdot 10^{18}$
eV$^{-1}$cm$^{-3}$ and the Drude conduction within the
crystallites $g(E_{\rm F}) = 5 \cdot 10^{21}$ eV$^{-1}$cm$^{-3}$
are clearly significant. The result is as anticipated from the
estimated charging energy of the order of 50 meV, and shows its
importance for the dc powder conductivity.

For Phthalcon-11 the very low number of carriers involved in the
hopping process is similar to the number of carriers that
determines the Drude contribution in the crystallites. Since the
mean size of the particles is 20 times larger than for ATO, the
charging energies will be of the order of 3 meV, and hence are
expected to be negligible at room temperature.

\subsection{Implications}

In polymer nanocomposites with building blocks formed by
diffusion-limited cluster aggregation, the airy structure of the
particle network gives a strong reduction in conductivity of the
composite compared to the filler (for the Phthalcon11/polymer
composite a factor $10^4$). \cite{Huijbregts06JPC} This effect can
be compensated by using better conducting particles. Particles of
ATO or ITO seem to be well-suited as the material is known to be
very well-conducting. In addition, ATO crystallites are relatively
easily obtained in sizes around 7 nm, and `when properly dispersed
can give polymer composites with a low percolation
threshold.\cite{Soloukhin07} However, even if the filler
nanoparticles in the composite touch, they will not be in better
contact than in a densely packed powder. As shown here for ATO,
for these small crystallites the DOS involved in $\sigma_{\rm dc}$
is dramatically reduced due to the shift of the energy levels away
from the Fermi level by Coulomb charging effects. As a
consequence, an additional four orders of magnitude in
$\sigma_{\rm dc}$ are lost compared to the bulk value.

Other systems where size restrictions are expected to be present
might be conveniently studied in a similar way. For example
several chalcogenide alloys exhibit a pronounced contrast between
the optical absorption in the metastable rocksalt after the
intense laser pulse and in the initial amorphous phase.
\cite{WelnicWuttig07} As shown by extended x-ray absorption fine
structure spectroscopy (EXAFS) the resistive change after the
intense laser recording pulse goes together with a crystallization
process, where also small domains are inherently present. Our
dielectric method might visualize to what extent the domain walls
after crystallization limit the conductivity and have consequences
for the band structure calculations. If the walls become real
barriers quantum size effects in the small domains will invalidate
the use of periodic boundary conditions in the calculations.  Also
the glassy behavior in the conductance of deposited indium-tin
oxide samples in the insulating regime,
\cite{BenChorin93Vaknin00Orlyanchik07} and of quench-condensed
insulating granular metals \cite{Kurzweil07} might be further
clarified by the use of our dielectric approach and analysis.
Scanning the frequency will reveal the evolution of the length
scales and DOS involved in the relaxation processes.

\section{Conclusions}

By combining data of sub-THz transmission with infrared
transmission and reflection we were able to explain the full
frequency response of densely-packed nanosized crystallites using
the parameters for Mott variable-range hopping, for
heterogeneity-induced enhanced ac response, for phonon- or
photon-assisted nearest-neighbor hopping, and for the Drude
response of individual nanocrystals. For 7 nm antimony-doped
tin-oxide particles the analysis unambiguously quantified the
reduction of the density of states involved in the dc conduction
compared to the value extrapolated from the Drude response at
infrared frequencies. Dielectric scans with a similar analysis
will also be revealing in other systems where size limitations are
expected to play a role.

\begin{acknowledgments}
It is a pleasure to acknowledge  Roel van de Belt of Nano Specials
(Geleen, The Netherlands), who made the ATO samples available, and
Matthias Wuttig from the Physikalisches Institut of the RWTH
Aachen University in Germany for fruitful discussions about
phase-change materials. This work forms part of the research
program of the Dutch Polymer Institute (DPI), project DPI435.
\end{acknowledgments}


\begin{thebibliography}{10}
\bibitem{noteCdSe1}
{The energy levels are discrete and assumed to have an overall
homogeneous  linewidth around 100 meV.}
\end{thebibliography}
\end{document}